\documentclass[sigconf]{acmart}
\acmConference[ICSE 2022]{The 44th International Conference on Software Engineering}{May 21–29, 2022}{Pittsburgh, PA, USA}
\usepackage{booktabs} 
\usepackage{multirow}
\usepackage{multicol}
\usepackage{enumerate}
\usepackage{tabularx} 
\usepackage{tabu}
\usepackage{threeparttable}
\usepackage{enumitem}
\usepackage{comment}
\usepackage{verbatim}
\usepackage[hyphenbreaks]{breakurl}
\usepackage{microtype}
\usepackage[utf8]{inputenc} 
\usepackage{balance}
\usepackage{siunitx}
\usepackage{tcolorbox}
\usepackage{graphicx} 
\usepackage{float} 
\usepackage{subfigure}
\usepackage{makecell}
\usepackage{diagbox}
\usepackage{soul}  
\usepackage{adjustbox}
\usepackage[figuresright]{rotating}

\newcommand{\tabincell}[2]{\begin{tabular}{@{}#1@{}}#2\end{tabular}}

\copyrightyear{2022}
\acmYear{2022}
\setcopyright{rightsretained}
\acmConference[ICSE '22]{44th International Conference on Software Engineering}{May 21--29, 2022}{Pittsburgh, PA, USA}
\acmBooktitle{44th International Conference on Software Engineering (ICSE '22), May 21--29, 2022, Pittsburgh, PA, USA}
\acmDOI{10.1145/3510003.3510205}
\acmISBN{978-1-4503-9221-1/22/05}

\AtBeginDocument{%
  \providecommand\BibTeX{{%
    \normalfont B\kern-0.5em{\scshape i\kern-0.25em b}\kern-0.8em\TeX}}}

\newtcolorbox{summarybox}[1][]
{
    sharp corners,
    left=1mm,
    right=1mm,
    boxrule=0.3mm,
    colback=yellow!30!white
    #1,
}

\begin{document}

\title{What Makes a Good Commit Message?}

\author{Yingchen Tian}\authornote{Yingchen Tian and Yuxia Zhang made equal contributions to this work.}
\affiliation{
  \institution{Beijing Institute of Technology}
  \city{Beijing}
  \country{China}
}
\email{tianyc10@foxmail.com}

\author{Yuxia Zhang}\authornotemark[1]\authornote{Corresponding authors}
\affiliation{
  \institution{Beijing Institute of Technology}
  \city{Beijing}
  \country{China}
}
\email{yuxiazh@bit.edu.cn}

\author{Klaas-Jan Stol}
\affiliation{
  \institution{University College Cork and Lero}
  \institution{School of Computer Science and IT}
  \city{Cork}
  \country{Ireland}
}
\email{k.stol@ucc.ie}

\author{Lin Jiang}
\affiliation{
  \institution{Beijing Institute of Technology}
  \city{Beijing}
  \country{China}
}
\email{jianglin17@bit.edu.cn}

\author{Hui Liu}\authornotemark[2]
\affiliation{%
  \institution{Beijing Institute of Technology}
  \city{Beijing}
  \country{China}
}
\email{liuhui08@bit.edu.cn}

\renewcommand{\shortauthors}{Tian et al.}

\begin{abstract}
A key issue in collaborative software development is communication among developers. 
One modality of communication is a commit message, in which developers describe the changes they make in a repository. As such, commit messages serve as an ``audit trail'' by which developers can understand how the source code of a project has changed---and why. Hence, the quality of commit messages 
affects the effectiveness of communication among developers.
Commit messages are often of poor quality as developers lack time and motivation to craft a good message.
Several automatic approaches have been proposed to generate commit messages. However, these are based on uncurated datasets including considerable proportions of poorly phrased commit messages. 
In this multi-method study, we first define what constitutes a ``good'' commit message, and then establish what proportion of commit messages lack information using a sample of almost 1,600 messages from five highly active open source projects. We find that an average of circa 44\% of messages could be improved, suggesting the use of uncurated datasets may be a major threat when commit message generators are trained with such data.  
We also observe that prior work has not considered semantics of commit messages, and there is surprisingly little guidance available for writing good commit messages. To that end, we develop a taxonomy based on recurring patterns in commit messages' expressions.   
Finally, we investigate whether ``good'' commit messages can be automatically identified; such automation could prompt developers to write better commit messages. 
\end{abstract}



\maketitle

\section{Introduction}\label{sec: intro}
\begin{figure}[!b]
\centering 
\includegraphics[width=7cm,height=2cm]{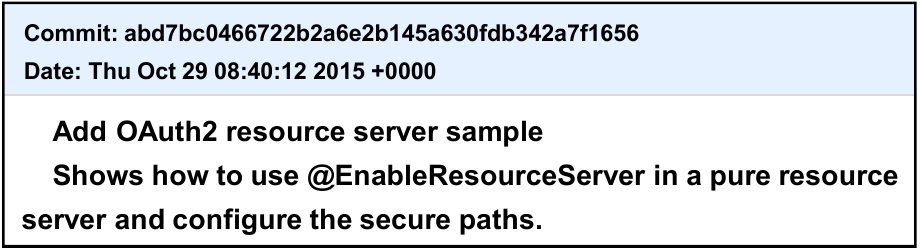} 
\caption{Example of a commit message from Spring-boot} 
\label{Fig.messageExample} 
\end{figure}

Collaborative software development is an inherently social activity, and is commonly facilitated through version control systems (VCS) such as Git \cite{zagalsky2015emergence, zhu2016effectiveness}. A VCS maintains a record of code changes, and manages simultaneous access to development artifacts \cite{zhu2016effectiveness}. 
Each commit should contain both changes to source code (and other stored artifacts) and a message that explains \textit{what} changes are made, and \textit{why} 
\cite{2010Automatically, mockus2000identifying}. Figure \ref{Fig.messageExample} shows an example of a commit message from the Spring-boot project \cite{springboot}. 
A well-written message is needed to communicate the context of a change to collaborators which allows them to review the change and understand its impact \cite{2012How, 2015agrawal}. 
For long-lived projects, such as the Linux kernel,  
commit messages might be the only source of information left for future developers to understand what changes were made and why those were made, when the original developers have left a project \cite{tan2019communicate}.

Software development is increasingly done in distributed settings involving developers from many different cultures and backgrounds. As well, in the past 20 years, commercial participation in open source software (OSS) projects has increased dramatically \cite{Zhang2017domination, zhang2019companies, zhang2020companies, Zhou2016Inflow}, leading to further diversification in the developer workforce on OSS projects that have become essential building blocks for many software organizations. This in turn may further diversify the quality of commit messages as individual developers and organizations may exhibit different development cultures and habits.

Several researchers have found that the quality of commit messages in repositories varies due to a lack of motivation or time \cite{5463344, Dyer2013, liu2018neural, liu2020atom}. For example, previous studies observed that ca. 14\% of commit messages in over 23,000 OSS projects were completely empty, 66\% of the messages contained only a few words, and only 10\% of commits had messages containing ``normal'' descriptive English sentences \cite{Dyer2013}.
Chahal and Saini \cite{chahal2018developer} proposed a syntactic model to calculate the quality of commit messages. However, this model can only assess the quality at the syntactic level through evaluation of ``rules,'' such as 
\textit{``the first character of the subject line should be capitalized''};
this model does not consider the semantics of commit message contents.
Further, developers may not know what kind of information should be written to produce a good commit message \cite{tan2019communicate}. The practitioner community is keen to help their contributors understand how to write a good commit message, as evidenced by guidelines such as: \textit{``the commit message should describe what changes our commit makes to the behavior of the code, not what changed in the code''} \cite{writing}.

To help developers write commit messages (addressing a potential lack of motivation and time), several tools have been proposed that can generate commit messages automatically \cite{2010Automatically, cortes2014automatically, liu2018neural}. Tools like DeltaDoc \cite{2010Automatically} and ChangeScribe \cite{cortes2014automatically} can produce detailed messages based on the changes contained within a commit, that can mainly answer \textit{what} was changed. 
However, the generated messages cannot reveal \textit{why} the change was necessary. Inspired by previous observations that commit messages follow certain patterns \cite{mockus2000identifying}, recent approaches \cite{2017Automatically, jiang2017towards, liu2018neural} generated commit messages from prior changes and their associated commit messages, which may contain the rationale for any code changes. For those approaches, the quality of generated commit messages relies on messages of similar code changes in the training data. A major problem is that the quality of commit messages in OSS projects used in training datasets for automatic generators might vary considerably. Therefore, using low-quality commit messages in training datasets introduces a major threat if these ``poor'' commit messages are not filtered out. Moreover, when the generated messages are the same or similar to the low-quality messages in testing datasets, automatic tools may yield an artificially high precision. Unfortunately, many models proposed thus far were trained and evaluated on datasets of commit messages, which simply removed trivial messages without paying attention to the content quality of these messages. 

This state of affairs leaves a number of important open questions unanswered.
First, (RQ1) 
to what extent do poorly composed commit messages exist? To the best of our knowledge, no prior work has defined or analyzed what makes a ``good'' commit message, and subsequently analyzed commit messages to assess the extent of how the quality of messages varies. 
Second, (RQ2) having established what makes a good commit message at a high level, what are recurring patterns of how these well-written messages are expressed? As we observed, the current state of the art does not consider the semantics of messages, only their syntax. 
Finally, future tools that could assist developers in writing good commit messages should be able to recognize whether a written message is ``good'' or not---for example, tools that can prompt developers in real-time as they attempt to make a commit may help improve the quality of commit messages. These tools would also be very useful for researchers in constructing high-quality datasets of commit message generation. Hence, our third research question is (RQ3): can commit messages of good quality be automatically identified? 

To answer these questions, we conducted a multi-method study. We first studied and analyzed a set of ca. 1,600 commit messages sampled from five major OSS projects. We defined a ``good'' commit message as one that explains what was changed, and why a change was made. We found that around 44\% of commit messages lack `why' or `what' information. 
This highlights the risk of generating messages based on an unfiltered training dataset that includes low-quality commit messages. 
Second, we qualitatively analyzed 252 messages with ``good'' message labels to identify expression characteristics. 
Third, we built a model based on Bi-LSTM to automatically identify well-written messages, achieving good performance.

This paper makes a number of practical and theoretical contributions to the literature on understanding and identification of high-quality commit messages. Specifically, we 1) propose a set of criteria for identifying well-written messages; 2) demonstrate that considerable proportions of commit messages lack essential information, thus highlighting that this variation in quality requires measures to mitigate; 3) propose a taxonomy of the expressions of commit messages; and 4) build an automated classifier to identify well-written messages.

In the remainder of this paper, we review related work in Sec. \ref{sec: relatedwork}, outline our multi-method research approach in Sec. \ref{sec: approach}, and present the results of our study in Sec. \ref{sec: results}. We discuss the implications for research and practice in Sec.~\ref{sec: discussion}. We present threats to the validity of our reported findings in Sec. \ref{sec: limitation} and conclude the paper in Sec.~\ref{sec: conclusion}.

\section{Related Work}\label{sec: relatedwork}
Commit messages constitute an important modality in collaborative software development for sharing knowledge among developers and in establishing an audit trail of the evolution of a software project. 
We discuss prior literature that has focused on understanding and utilizing commit messages and how to automatically generate commit messages.

Commit messages are a key resource when addressing several software engineering challenges. 
One stream of research has focused on classifying code changes into different types by utilizing commit messages manually or automatically to assist maintenance \cite{dos2020commit, purushothaman2005toward, mockus2000identifying}. For example, Mockus and Votta \cite{mockus2000identifying} identified three types of commits: adaptive, corrective, and perfective, consistent with Swanson's typology of maintenance activities \cite{swanson1976dimensions}. 
Based on the proposed commit types, numerous classification models have been proposed, and commit messages play an important role \cite{dos2020commit, yan2019characterizing, levin2017boosting}.  

A second stream of research has focused on the measurement of quality of code changes by analyzing commit messages. For example, Agrawal et al. \cite{2015agrawal} studied the evolution of commit quality in five projects by measuring (among others) the number of unique commit messages, and found that the quality of commits declined over time. Santos et al. \cite{2016santos} studied the relationship between ``unusual messages'' and code quality in commits, and found that unusual messages correlate with build failures, suggesting that these messages serve as a warning sign. 

While it is clear that commit messages play an important role in communication among developers, developers may lack time or motivation to craft good commit messages that clearly communicate what is being committed.
To address this, several scholars have proposed automatic approaches to automatically generate messages. Some of them are rule-based or use predefined templates \cite{2010Automatically, cortes2014automatically, 2015changescribe, 2016shen}. For instance, Buse and Weimer \cite{2010Automatically} used symbolic execution to generate path predicates between versions of code changes, then populated pre-defined templates and applied summarization transformations to generate commit messages for code changes. Two important limitations of these commit messages generated based on templates are (1) a lack of flexibility, and (2) they cannot convey the intent of committing changes, which only exists in a developer's mind until it is written. Recent studies rely on advanced techniques, such as information retrieval and deep learning \cite{huang2020learning, 2017Automatically, liu2020atom, xu2019commit, nie2021coregen, liu2018neural,2018loyola} to generate commit messages automatically. These tend to rely on reusing messages of similar code changes. For example, Huang et al. \cite{2017huang} calculated syntax, semantic, pre-syntax, and pre-semantic similarities of changed code fragments between two versions to find similar code changes and reuse their messages. While specific models vary in their techniques, a common feature is that they take prior commit messages as a key input.
For these information retrieval and deep learning based tools, the quality of the manually written commit message is difficult to guarantee \cite{Dyer2013, liu2018neural}, which may threaten the effectiveness of these tools.

A few studies investigated the content of commit messages but mainly focused on specific aspects. For example, Alomar et al. explored how developers document their refactoring activities in commit messages, and found that developers tend to explicitly mention the improvement of certain quality attributes and code smells \cite{alomar2019can}. Chahal and Saini \cite{chahal2018developer} constructed a model that can judge the quality of commit messages by calculating 11 syntactical measures. Text content in other OSS development activities has been studied, such as what/how to document when submitting patches \cite{tan2019communicate} and what information is needed in a bug report \cite{zimmermann2010makes}. The results of our study on the distribution and expression categories of good commit messages and their relationship with maintenance activities can complement prior understanding of what is a good commit message and how to write one. Moreover, we propose a good-message identification tool that can be used to prompt developers to write better commit messages and build high-quality datasets for the task of automatically generating commit messages.

\section{Study Design}\label{sec: approach}
\begin{figure*}[!t]
\centering 
\includegraphics[width=15cm]{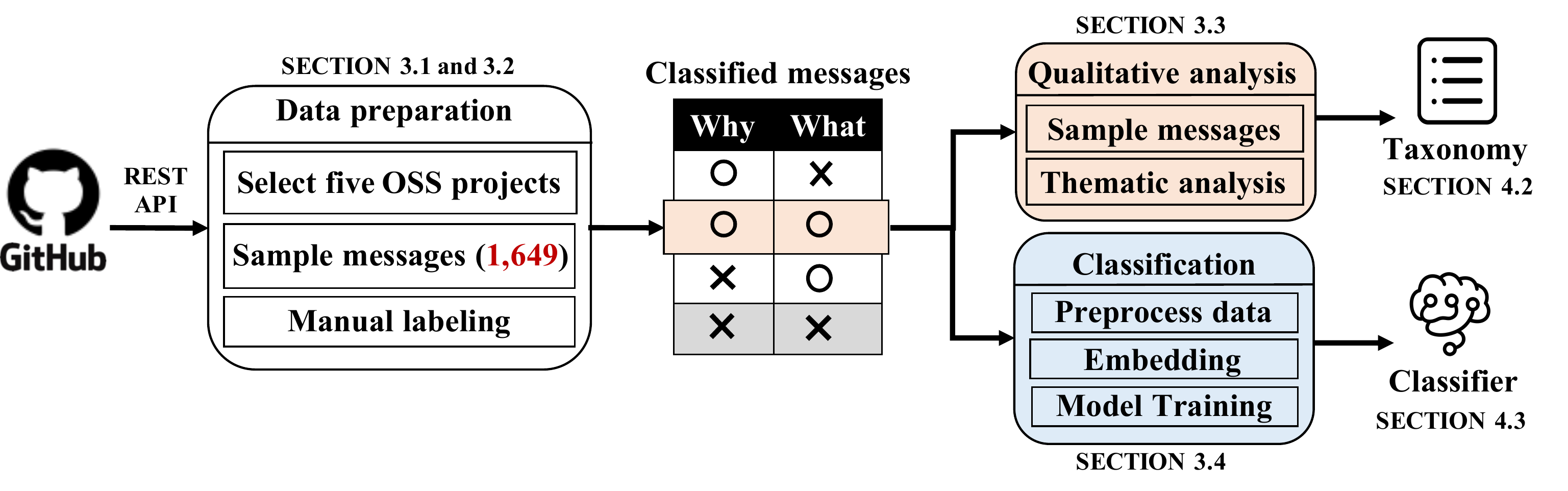} 
\caption{Overview of our approach} 
\label{Fig.overview} 
\end{figure*}

To address the three research questions introduced in Sec. \ref{sec: intro}, we conducted a multi-method study. To address RQ1, we compiled a dataset of commits and manually classified the messages based on our definition of a ``good'' commit message. Sec. \ref{sec: data} describes sample selection, data collection, and data processing steps. Sec. \ref{sec: identify} describes our approach to classifying commit messages. To address RQ2, we develop a taxonomy that describes how commit messages convey ``what'' and ``why'' information  (see Sec. \ref{sec: characterizing}). Finally, to address RQ3 we propose a model that could identify these well-written messages automatically (see Sec. \ref{sec: model}). Figure \ref{Fig.overview} presents an overview of this approach. The appendix offers a replication package \cite{onlineAppendix}.

\subsection{Data Collection and Preprocessing}\label{sec: data}
To ensure that our sample would contain sufficient high-quality commit messages, we selected active and popular projects with a high level of collaboration. 
We assumed that those projects would have at least some non-trivial portion of good commit messages. Considering the impact of different programming languages on software development \cite{10.1145/3340571}, in this study, we focused on projects written in Java, one of the most popular programming languages in GitHub \cite{javaPopular20210net} and widely used in industry. We selected the top 100 Java projects sorted by their ``star'' rating in GitHub. While reviewing these projects, we prioritized projects that had previously been subject of studies that focused on automatic generation of commit messages. By addressing some of the limitations of those studies, we seek to offer the results of this study in future improvement of those prior studies focusing on commit message generation. As a result, we selected five OSS projects for this study, listed in Table~\ref{summaryOfProjects}. \textbf{\textit{Spring-boot}} \cite{springboot} helps developers create and run Spring-based applications with less configuration. \textbf{\textit{Apache Dubbo}} \cite{dubbo} is a high-performance micro-service development framework.\textbf{\textit{ Okhttp}} \cite{okhttp} is a HTTP client.
\textbf{\textit{Junit4}} \cite{junit4} provides the ability to write repeatable unit tests. \textbf{\textit{Retrofit}} \cite{retrofit} is a type-safe HTTP client for Android and Java. Each of the five widely investigated \cite{2015changescribe, yan2019characterizing, chen2019code, di2018preliminary, aman2017survival, zampetti2017open} projects has more than 150 contributors, over 8,000 stars, and thousands of commits, indicating that there is active collaboration happening within these projects.

We collected all the commits from the five projects using GitHub's REST API \cite{githubrestapi} up to February 2021. 
In the data collection, we only considered commits in which the messages are written in English. This resulted in a dataset containing 41,886 commits (see Table~\ref{summaryOfProjects}). We eliminated commit messages generated automatically by tools (bot messages) based on fixed patterns because this study focuses on messages written manually. Based on the patterns identified by existing work \cite{erlenhov2019current, dey2020exploratory, Amreen2020, dey2020detecting}, these bot messages can be easily identified and filtered. Table~\ref{patternOfBotMsg} shows several patterns of bot messages in our dataset, which were excluded (ignoring cases). After this data cleaning step, 29,348 commit messages remained.

\begin{table}[!h] 
\caption{Dataset summary statistics (until Feb 2021)}
\sisetup{
    group-digits=true,
    group-minimum-digits=4,
    mode=text,
    detect-weight=true, 
    detect-family=true,
    group-separator={,}
}

\begin{tabular}{cccS[table-format=7.0]S[table-format=7.0]}    
    \toprule    Project & Start & \#Contrib. & {\#Commits} & {\#Cleaned}\\    
    \midrule \hangindent=3mm Spring-boot & Oct-2012 & 812 & 30072 & 21169 \\
        Apache Dubbo & Jun-2012 & 404 & 2687 & 2249\\ 
        Okhttp & Jul-2012  & 236 & 4800 & 2817\\ 
        Junit4 & Apr-2009 & 151 & 2467 & 2035\\ 
        Retrofit & Sep-2010 & 152 & 1860 & 1078\\
        \midrule
         & &  &41886 & 29348\\
    \bottomrule   
    \end{tabular} 
\label{summaryOfProjects}
\end{table}

\begin{table}[!h]
\caption{Non-human written message patterns}
    \begin{tabular}{p{0.3cm}p{7.2cm}}
        \toprule 
        No. & Pattern\\
        \midrule 
        1 & merge branch <branch> (of <project url>) (into <branch>) \\ 
        
        2&\tabincell{l}{merge remote-tracking branch <branch> (into <branch>)} \\ 

         3&[maven-release-plugin]\\

         4&\tabincell{l}{...cherry picked from commit <commit url>}\\

         5&\tabincell{l}{Next development version <version number>}\\

         6& message written by non-human accounts, such as Spring Operator, dependabot[bot], no author\\

        \bottomrule 
    \end{tabular} 
    
    \label{patternOfBotMsg}
    \begin{tablenotes}
        \footnotesize
        \item[*] ``<branch>'' means branch name of the project, ``(...)'' means optional
    \end{tablenotes}
\end{table}

\subsection{Identifying Well-Written Messages} \label{sec: identify}
It is known that the quality of commit messages varies \cite{5463344, Dyer2013, liu2018neural}, but a widely recognized standard of high-quality messages is as of yet lacking. Before investigating the characteristics of well-written messages, we constructed the standards to identify them via a survey of both academic papers and developer forums and validate the standards with experienced OSS developers, as described below:
\begin{itemize}
    \item To obtain a scientific perspective of commit messages, we identified and reviewed 46 relevant studies (the full papers are listed in our online appendix \cite{onlineAppendix}),  mainly focusing on the expectation of commit messages and whether there is a standard of good messages.
    \item To obtain a pragmatic and practice-oriented view of commit messages, we used Google's search engine with the phrase ``good commit message.'' In the top 50 results sorted by relevance, we manually selected and studied the online records (and their references) from OSS communities or OSS developers (the whole set of links is included in the appendix \cite{onlineAppendix}). Furthermore, we also solicited opinions from 30 experienced OSS developers to gather their views on what constitutes a good message; we defined ``experienced'' here as those developers who contributed more than 10 commits to the studied projects. 
\end{itemize}

Through qualitative analysis of the selected records, we observed that the most frequently recognized expectation of a commit message is \textit{\textbf{to summarize the changes in this commit (noted as `What') and describe the reasons for the changes (noted as `Why').}} 
The surveyed developers showed a high degree of consistency in the content of commit messages: approximately 93\% of them held the view that a commit message should summarize \textit{what} was changed, and describe \textit{why} those changes are needed. 
That is, rather than only summarizing changes or describing the reasons for the changes, a commit message should have both What and Why information to help collaborators understand the changes.

Therefore, we conducted this research based on the hypothesis that a good commit message should contain a justification (i.e., \textbf{``Why''}) that describes the motivation of the change, and a change summary (i.e., \textbf{``What''}). 
Depending on whether or not the two key elements, Why and What, are included, we divided the commit messages in our dataset into four types: \textit{``Why and What''} (containing both), \textit{``No Why''} (only What information, but no Why); \textit{``No What''} (only Why, but no What); and \textit{``Neither Why nor What''.} 
We note that Why information for certain changes might be common sense. For example, the commit message \texttt{``fix typo a->an''} omits the reason of making this commit, which can be easily inferred as ``improving readability.''
From the perspective of reducing developers' workload, we did not classify these commit messages as ``No Why'' because the rationale is trivial. Similarly, some What information could be easily inferred from 
\texttt{diffs},\footnote{\texttt{Diffs} are raw content of changes generated using the \textit{git diff} command to show differences between different versions of commits.} and we took the same approach as we did for Why. More details of the categories of common sense Why and easy-to-infer What can be found in Sections~\ref{sec: why patterns} and \ref{sec: what patterns}. 
Further, we found some commits express Why information by providing a link to an issue report or pull request, which usually includes a detailed motivation and discussions of the change \cite{zhu2016effectiveness, Anvik2006}. 
This approach is controversial, however: one school of thought argues that these linked resources provide a convenient way to offer full details that lay out the rationale for a change \cite{sawant2018features, tan2019communicate}. 
Another school of thought argues that such links pose a risk as they might go stale, resulting in a loss of the information they point to, thus resulting in commit messages that are difficult to understand \cite{le2015rclinker}.
In this study, we took the former view and treated links of issue reports and pull requests in commit messages as a way to provide Why information.

To study the distribution of the four message types in the selected projects (Table~\ref{summaryOfProjects}), we used the clustered random sampling technique \cite{2016A} on the five projects' commit data and selected 1,649 commit messages (confidence level: 95\%, margin of error: 5\%), which were committed by 339 developers. On average, each developer submitted approximately five commits in our dataset.  
The first two authors of this study labeled the commit messages independently, categorizing each into one of the four message types. During the labeling process, we identified and eliminated 52 non-atomic commits, where more than one change was submitted. 
Making multiple changes in a single commit is considered bad practice as it may reduce maintainability \cite{gitbook, git-commit, 2015agrawal}. 
After labeling the messages (1,649 minus 52 that were removed), Cohen's kappa coefficient of agreement \cite{cohen1960coefficient} between the two authors was 0.91. 
As for the messages labeled differently, we held several meetings to resolve 66 (approx. 4.1\%) disagreements. If the first two authors failed to reach an agreement on the type, a third author acted as an arbitrator.  
Moreover, we validated the labeled results by conducting a survey with OSS contributors. Specifically, we selected 958 experienced contributors (i.e., those who contributed more than ten commits) from the top 100 Java projects sorted by their star count and sent them a questionnaire (see the appendix \cite{onlineAppendix}) to solicit their views. 
We received 30 valid responses (a response rate of 3.1\%). After calculating the 120 results labeled by developers (each respondent labeled four commit messages), we found that 102 results were labeled consistently with ours. This indicates that the accuracy of the dataset reached almost 85\%.

\subsection{Characterizing Well-Written Messages} \label{sec: characterizing}
In the second phase, we sought to identify the characteristics of well-written messages (labeled in Sec.~\ref{sec: identify}). 
We manually sampled 271 (confidence level: 95\%, margin of error: 5\%) commit messages with a labeled type as \textit{``Why and What''}---that is, messages that contained both Why and What information, and thus did not miss important information. Among the 271 commits, we removed 19 commits that only use links of pull requests or issue reports to express Why. 
For the remaining 252 messages, we used thematic analysis \cite{cruzes2011recommended} to characterize how developers express Why and What information in the commit messages, according to the following process.
(1) We first read and analyzed all the commit messages, to understand how developers described code changes and motivation, and identified phrases that expressed Why and What. (2) We reread the whole commit messages and related phrases carefully to generate initial codes and organize them in a systematic way. (3) After completing the generation of the initial code, we aggregated codes with similar meanings, and identified an initial theme representing that cluster. After this step, all codes were divided into one of the initial themes, which helped in identifying any emergent patterns that characterized the descriptions of Why and What.
(4) We then reviewed the initial set of themes to identify opportunities to merge similar themes. By clarifying the essence of each theme, similar themes were merged into a new theme, or a theme was included as a sub-theme.
(5) In the last step, we defined the final set of themes. 

To reduce any researcher bias, steps (1) to (4) described above were performed independently by the first two authors \cite{runeson2009guidelines}. After this, a sequence of meetings was held to resolve conflicts and assign the final themes (step 5).

During thematic analysis of the 252 commit messages, we found the way developers describe Why and What in messages tends to vary across different types of maintenance activities. Therefore, we also classified commit types to investigate the relationship between message expression categories and maintenance activities. Prior literature provides a variety of code change classifications \cite{swanson1976dimensions, mockus2000identifying, yan2016automatically, rezk2021ghost}, but no consensus was reached regarding the different types of classification to which a commit refers. Therefore, after an ad-hoc literature review, we adopted the widely used definition of Mockus and Votta \cite{mockus2000identifying} for three commit types:
(1) corrective changes address processing, performance, and implementation failures; (2) adaptive changes represent changes in the data environment or processing environment. For example, to implement a new function; and (3) perfective changes, which focus on improving non-functional attributes such as efficiency, performance, cleanup, etc. Then, we identified commit type by deductive thematic analysis \cite{pearse2019illustration}, which can match the data with themes from extant research. More specifically, the first two authors independently took the theoretical propositions derived from Mockus and Votta \cite{mockus2000identifying} as a point of departure, and applied them to the 252 commit messages. We obtained a high level of consistency (Cohen's kappa coefficient = 0.92) between the two coders. The two coders discussed and resolved any disagreements.

\subsection{Automatic Identification}\label{sec: model}
In the third phase, we sought to develop a solution that could automatically identify well-written messages. As we defined high-quality commit messages as those containing both Why and What information, we first designed two classifiers that could automatically identify whether a message contains Why (labeled \textit{\textbf{C-Why}}, and \textit{\textbf{C}} means ``classifier'') and What (labeled \textit{\textbf{C-What}}) separately. Training the two separate classifiers can offer more fine-grained feedback to developers by indicating which of the two key elements (Why and What) is missing.
We then selected and combined the two classifiers with the best performance to automatically identify well-written messages that contain both Why and What (labeled as \textit{\textbf{C-Good}}).

\subsubsection{Data Preparation.} We used the commit messages labeled in Sec.~\ref{sec: identify} to train and test the three classifiers. Commit messages usually include several tokens that are not ``natural language,'' such as links to pull requests. Since their full semantics are highly specific to the contents of the commits, which we do not consider in this paper, we replaced these tokens with placeholders indicating the kind of information, to ensure the models were not affected by such trivial commit content. Specifically, we identified and replaced the following tokens of non-natural language:
1) we replaced any URLs in a message with \texttt{``<X url>''}, where ``X'' refers to the types of URLs, i.e., ``pr'' (indicating pull request links), ``issue'' (indicating links of issue reports), and ``other.'' 
2) We replaced code elements in the messages with \texttt{``<X name>''}, where ``X'' refers to the types of code elements, such as method and file. These code elements were identified by comparing messages with the corresponding code changes. 3) We retained the paragraph information for commit messages by replacing newline characters with \texttt{``<enter>''}.

\subsubsection{Identification.}
It is likely that the original dataset is imbalanced, i.e., the number of messages that contain Why (or What) is larger than the messages that do not contain this information.
However, imbalanced datasets can cause machine learning (ML) models to focus on major categories and undervalue other minor categories \cite{AkbaniKJ04} that we are more concerned about in this paper. 
ML-based classifiers often use over-sampling methods to solve the data imbalance problem so as to achieve better performance. 
To this end, we tried three widely used over-sampling techniques, including random sampling with replacement \cite{japkowicz2000learning}, synthetic minority oversampling technique (SMOTE) \cite{chawla2002smote}, and the adaptive synthetic (ADASYN) \cite{he2008adasyn}, to prepare the data before training the classifiers, and selected the technique that produced the highest accuracy in each classifier.

Next, the input textual messages were vectorized.
The vectorization method, i.e., Bidirectional Encoder Representations from Transformers (BERT) \cite{devlin2018bert}, has been shown to exhibit good performance in natural language processing tasks including text classification \cite{gao2018neural, adhikari2019docbert}. 
We used BERT to embed the tokenized and padded commit messages and convert each message into a numeric vector.
A large number of techniques have been proposed to solve automatic classification tasks \cite{menard2002applied}.
We considered the most widely-used techniques to classify commit messages, including Long Short-Term Memory (LSTM) \cite{hochreiter1997long}, bidirectional Long Short-Term Memory (Bi-LSTM) \cite{schuster1997bidirectional}, Multi-Layer Perceptron (MLP) \cite{pal1992multilayer}, Logistic Regression \cite{menard2002applied}, Random Forest \cite{breiman2001random}, 
K-Nearest Neighbors (KNN) \cite{cover1967nearest}, Gradient Boosting Machine \cite{friedman2001greedy}, and Decision Tree \cite{quinlan1986induction}.
We evaluated their performance in our classification task, and selected the best performing approach.

\section{Results}\label{sec: results}
We now present the results of our study, addressing the three research questions outlined in Sec. \ref{sec: intro}. Sec. \ref{sec: rq1} presents the distribution of different types of commit messages. Sec. \ref{sec: rq2} presents a taxonomy of  well-written messages, comprised of expression categories that developers use to describe What and Why information. 
Sec.~\ref{sec: rq3} presents the performance results of our automatic classifier of well-written messages. To facilitate traceability, we provide quote messages, and provide identifiers in our database \cite{onlineAppendix}.

\subsection{Quality Distribution of Commit Messages}
\label{sec: rq1}
\begin{figure} 
\centering 
\includegraphics[width=8.5cm]{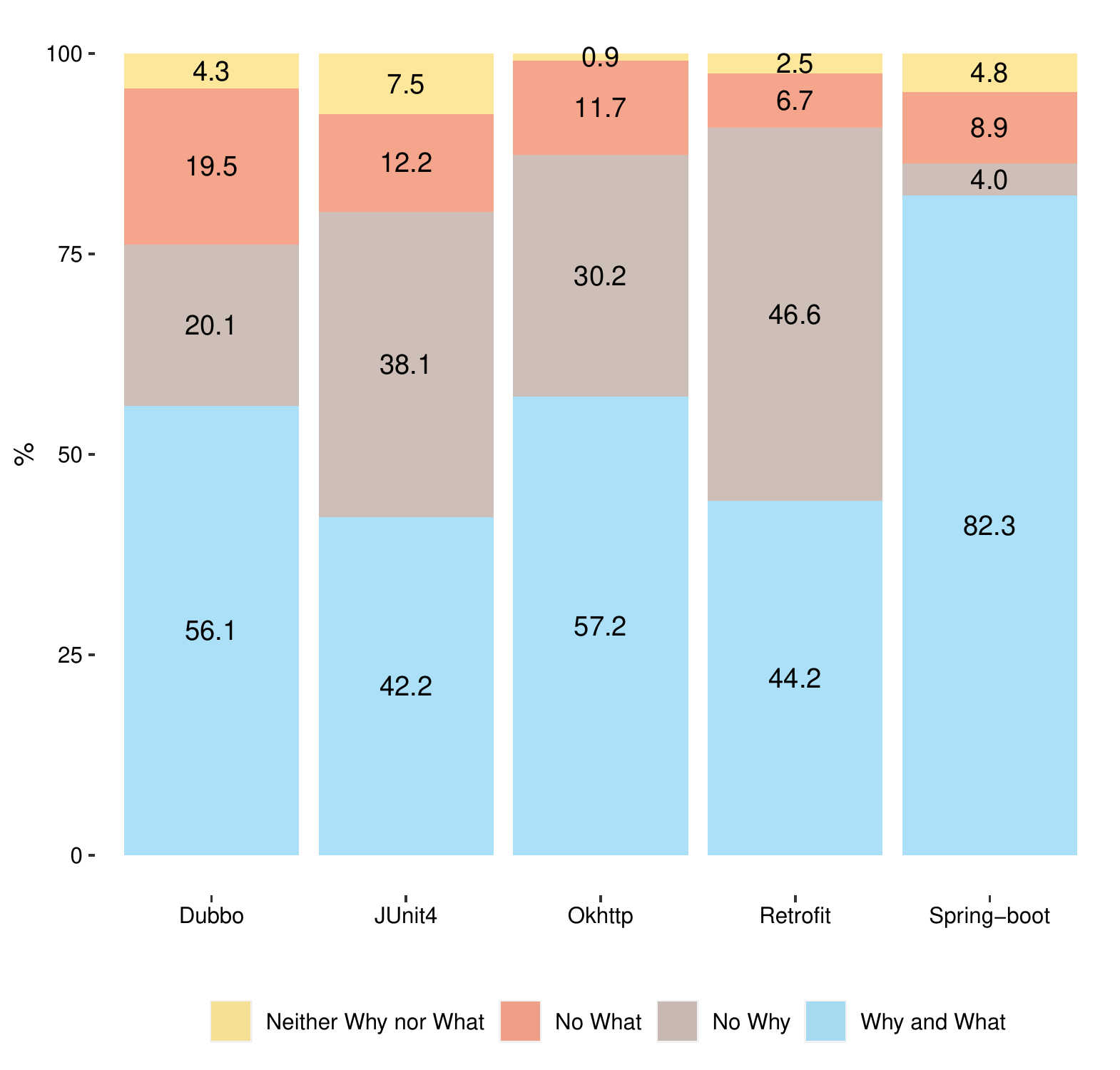}
\caption{Distribution of the four types of commit messages} 
\label{Fig: distribution} 
\end{figure}

We manually classified 1,597 commits (sampled from five OSS projects) into four types based on whether their messages contain Why and What information (see Sec.~\ref{sec: identify}). We calculated the distribution of these four types of commit messages in the five OSS projects. Figure~\ref{Fig: distribution} presents the results. We can see that the dominating type in four projects (except Retrofit) is well-written messages, i.e., these messages contain both Why and What information. The ratio of this message type in the five projects varies from ca. 42\% to ca. 82\%, with an average ratio of ca. 56\%, suggesting that around 44\% of commit messages have quality issues. This, in turn, suggests that any automated approaches to generate commit messages which are trained using datasets containing such large portions may be compromised, as the generated messages may have learned from such incomplete messages. While our sample of five projects is clearly not representative of the larger corpus of Java projects on GitHub, this finding does highlight a potential issue in terms of the effectiveness of existing tools.

As for the three questionable types of messages, ``No Why'' (containing only What information) accounts for the largest proportion with an average of 28\%, and is approximately twice the ratio (12\% on average) of the messages containing only Why information. It may indicate that writing the reasons of code changes is more challenging than describing what was changed. Further, the widely different proportions of ``No Why'' and ``No What'' may explain why generating Why information for code changes is harder than generating What information, which previous studies have borne out \cite{2018loyola, liu2018neural, liu2020atom}.

The type of messages that contain neither Why nor What accounts for the smallest proportion, ranging from ca. 1\% to ca. 8\%. To further investigate what information the type ``Neither Why nor What'' exactly contains, we manually analyzed all the 65 messages of this type. Following the same set of steps of thematic analysis \cite{cruzes2011recommended} described in Sec. \ref{sec: characterizing}, we identified five categories, each with unique characteristics:

\begin{itemize}
\item\textbf{Single-word message}: containing only one token which hardly expresses any information. ``\textit{Merge}'', ``\textit{Polish}'', and ``\textit{<file name>}'' are typical examples in our labeled data.
\item\textbf{Submit-centered message}: expressing nothing but the fact that it is a commit. For example, 
``\textit{Loader changes}.'' It is obvious that this is a commit, but it is impossible to know what the changes are, and why they were made.
\item\textbf{Scope-centered message}: explaining nothing but the scope of the changes. Typical messages of this category include ``\textit{minor changes in test}.'' Most of such messages contain qualifying words like ``major'' and ``minor'' and cannot express other information.
\item\textbf{Redundant message}: describing content that is easy to infer from code diffs. For example, ``\textit{Add <file name>}'', ``\textit{Delete <file name>}.'' Even without reading this message, the fact that a file was added or removed can be easily established.
\item\textbf{Irrelevant message}: describing something irrelevant to the change. A message such as \textit{``Kent \& Erich patch swallowing in Merlin''} is written to commit a non-empty message, which conveys no information at all.
\end{itemize}

The analysis above clearly demonstrates that a considerable portion of commit messages is problematic, i.e., they do not contain important information. That is, they either miss What or Why information, or both. As we discussed in Sec. \ref{sec: relatedwork}, several message generators have been trained and evaluated using commit messages as datasets, without filtering out sub-optimal messages. Consequently, a major threat was introduced due to using sub-optimal data. 
The fourth type, ``Neither Why nor What'', might be easy to identify and remove, but this type only accounts for a limited proportion. In other words, filtering this type of message cannot address the threat sufficiently, and a more powerful automatic identifier of good messages is needed.

\vspace{2mm}
\begin{summarybox}
{
\textbf{Summary for RQ1:} 
The quality of commit messages varies in the five studied OSS projects, with on average ca. 44\% of messages in need of improvement. Further, we identified five categories of messages that contain neither ``Why'' nor ``What.'' 
}
\end{summarybox}

\subsection{A Taxonomy of Commit Messages} 
\label{sec: rq2}
We now turn to RQ2, which seeks to shed light on what makes a well-written commit message. We identified the various categories of rationales (the ``Why'') as well as the contents (the ``What''), and analyzed the relationships between these categories and the typology of Mockus and Votta \cite{mockus2000identifying}. Note that multiple categories of Why (or What) may exist in the same message. Specifically, we 
introduce these categories in Sec. \ref{sec: why patterns} and \ref{sec: what patterns}, and present the prevalence of these categories in Sec. \ref{sec: link}. We summarize the detailed codes and statistics of the Why and What categories in our online appendix \cite{onlineAppendix}.

\subsubsection{Why Expression Categories}\label{sec: why patterns}
By analyzing the various ways in which developers express the change rationale (i.e., Why), we found that developers tend to express this directly or indirectly, or sometimes not at all when the reason for a change is common sense or can be explained by the change itself. 
Using the procedure outlined in Sec.~\ref{sec: characterizing}, we identified five main categories with 18 subcategories. We introduce these (sub)categories with examples as follows, including their counts and percentages (indicated in parentheses) in the 252 analyzed commit messages.

\paragraph{\textbf{Describe Issue}}
This category directly elaborates the motivation of a code change; it is mainly concerned with an issue in the current code implementation. Developers explained their motivation for a change by describing the issues and the specific scenario in which they occurred. This makes the context of a commit easier to understand for other contributors. We identified three subcategories:

\begin{itemize}

\item[\textsc{di1}] \textbf{Describe error scenario} (\#50, 19.8\%): this subcategory directly describes where and how an error occurs. It is the most common way of describing Why a change was made. Developers frequently indicated the source of a bug, or specified the steps for reproducing it, and also explained their impact; for example:
\textit{``As-is it throws unchecked exceptions on unexpected charsets. This is a problem because it can cause a misbehaving webserver to crash the client.''} [\#S243]

\item[\textsc{di2}] \textbf{Introduce issue report} (\#11, 4.4\%): this subcategory describes issues mainly by citing errors/defects or warnings from quality assurance tools. Some recognized or common tools are usually chosen to achieve the contributor's common understanding of the mistakes. For example, \textit{``remove warnings found by errorprone. [...] CallTest.java:2056: warning: [UnnecessaryParentheses] Unnecessary use of grouping parenthese [...]''} [\#S88]. This message cites a warning message from the tool ``errorprone'' \cite{error-prone} to describe the motivation for this commit.

\item[\textsc{di3}] \textbf{Describe shortcoming} (\#9, 3.6\%): this subcategory highlights the shortcomings or weaknesses in the current implementation, which is the motivation to make this commit. For example, \textit{``I'm unhappy with java.io: No timeouts [...] Features like mark/reset and available() are clumsy [...]''} [\#S141].

\end{itemize}

\paragraph{\textbf{Illustrate Requirement}} 
The second category, \textit{Illustrate Requirement}, describes the source of requirements that led to this commit. These requirements include the need for software development and addressing problems in the process of software maintenance. We identified three subcategories:

\begin{itemize}

\item[\textsc{ir1}] \textbf{Usage need} (\#11, 4.4\%): these commit messages describe specific needs or requirements of users in software development. This message helps other contributors understand the background and necessity of this change. For example, \textit{``Error-prone only works on pre-12 at the moment and we need this configuration to apply for all JDKs''} [\#S225].

\sloppy
\item[\textsc{ir2}] \textbf{Out of date} (\#24, 9.5\%): these commit messages indicate the obsolescence of some features or code. This includes the deprecation and subsequent removal of unused code objects such as classes, methods, or attributes as the software evolved. Other object version upgrades may also cause a dependency to become obsolete requiring modifications. For example: \textit{``Remove outdated key. The `spring.metrics.export.redis.aggregate-key-pattern' is no longer defined but was still referenced in the documentation.''} [\#S127].

\fussy 
\newdimen\origiwspc%
\newdimen\origiwstr%
\origiwspc=\fontdimen2\font
\origiwstr=\fontdimen3\font
  
\fontdimen2\font=1em
\item[\textsc{ir3}] \textbf{Runtime or development environment change} (\#11, 4.4\%):
\fontdimen2\font=\origiwspc
commit messages in this subcategory indicate an adaptation to the current code development or runtime environment. 
This includes changes to the implementation to adapt dependent functional changes, modify documents, return values, or examples to accommodate API modifications, etc. 
For example, \textit{``API has changed, fixing the example,''} indicates that the developer changed the example to match the changes of an API [\#S142].
\end{itemize}

\paragraph{\textbf{Describe Objective}} 
The third category of rationale provides the purpose of a change, such as the future prevention of defects or optimization of functionality or performance. 
We identified two types of objectives:
\begin{itemize}

\item[\textsc{do1}] \textbf{To fix defects} (\#9, 3.6\%): 
these commit messages make explicit that the purpose of a change is to resolve a defect. Different from \texttt{Describe error scenario} (\textsc{di1}) that describes how a defect occurs or can be observed, this type of expression clarifies how the proposed change will \textit{resolve} a defect, as in this example: \textit{``Fix concurrent problem of zookeeper configcenter, wait to start until cache being fully populated''} [\#S250].

\item[\textsc{do2}] \textbf{To make improvements} (\#13, 5.2\%): 
a message in this subcategory directly describes the improvement of an author's code implementation, e.g., functional improvement or non-functional goals. The message explains the reasons for the change by describing a specific promotion goal. Commit messages in this category usually use phrases such as \textit{to do something} or \textit{for something} to describe what an author seeks to achieve. For example, the message \textit{``AndroidLog: Added [...] methods for easier subclassing.''}[\#S181] clarifies that the change will simplify the act of subclassing.

\end{itemize}

\paragraph{\textbf{Imply Necessity}}
Different from commit messages in the previous three categories, messages in the \textit{Imply Necessity} category only \textit{indirectly} describe the need for changes. 
Developers indirectly described the necessity of changes using the messages that fall into the following subcategories:

\begin{itemize}
\item[\textsc{in1}] \textbf{Conventions and standards} (\#15, 6.0\%):
this subcategory describes or refers to any conventions or standards that are the basis for a change, thus demonstrating a sound rationale of the change.
Conventions are agreements among developers within the project, or could also be industry-wide conventions. Standards are written specifications, often rather technical and more formal, and thus are more rigid than conventions. 
A common understanding among contributors expressed in commit message makes a commit easier to understand. For example, one developer referred to a convention regarding the location of tests, thus explaining why the commit moves the location of the test: 
\textit{``it is common to add tests to the same package as the class under test''} [\#S8].

\item[\textsc{in2}] \textbf{Relation to prior commits} (\#9, 3.6\%): 
these commit messages explain the relationship between the current commit and any commits that were already merged into the repository. These messages clarify the motivation by improving any problems with a prior accepted change or using the new features introduced by a previous change, etc. Therefore, while the \texttt{Describe Issue} category may be used to indicate that there is a problem \textit{before} changing it, the nature of this subcategory is to introduce prior accepted changes as the context for the current commit. For example, this commit adds a test because \textit{``the code was changed by commit <other url> but unfortunately the test was not part of the commit''} [\#S40].

\item[\textsc{in3}] \textbf{Relation to an implemented feature} (\#9, 3.6\%): 
some commits are related to an existing feature, and this relationship provides the context of the new commit. The current change is part of a large operation that may be underway, such as the message [\#S44] indicates that the current change is \textit{``a short step on the road to HTTP body format agnostic support.''} The current change may also be preparation work for an accepted feature, such as message [\#S41] which directly explains that the change will \textit{``make a future change easier to land.''} 
In the context of established goals (achieving functionality or larger operations), the motivation for this change will be clear.

\item[\textsc{in4}] \textbf{Improvements and benefits} (\#39, 15.5\%): 
the last subcategory refers to commit messages that indirectly describe the need for a change by explaining the improvements and benefits that the change will bring. Such commits may include either functional or non-functional improvements such as readability and maintainability. The commit message may also include a comparison between ``before'' and ``after'' the commit, which gives collaborators a better understanding of the motivation for change. For example, the following commit message suggests a proposed improvement and the associated benefit: \textit{``Use custom exception type [...]. Since we omit the stack trace, this more clearly indicates the source being from Retrofit's mock behavior''}[\#S103].
\end{itemize}

\paragraph{\textbf{Missing Why}}
This category includes commit messages that do not offer a rationale, for example, when it is common sense or easy to infer. In such cases, there is no need to provide a rationale. These include the following six subcategories:

\begin{itemize}
    
    \item[\textsc{mw1}] \textbf{Test cases} (\#4, 1.6\%): 
    these commits involve adding test cases to the repository; in many projects, there is consensus among developers to add test cases for each feature. For example, \textit{``tests for canceling async requests.''} [\#S21]. 
    
    \item[\textsc{mw2}] \textbf{Typographic fixes} (\#7, 2.8\%): 
    these commits involve the correction of typographic errors. Fixing such errors helps to increase the correctness and readability of code or documentation, for example:
    \textit{``fix typo a->an''} [\#S4].

    \item[\textsc{mw3}] \textbf{Text file changes} (\#13, 5.2\%): 
    these commits involve changes made only in text files. Such files have specific functions, such as ``ChangeLog'' files that record changes, ``README'' files that outline the project, and so on.
    
    \item[\textsc{mw4}] \textbf{Annotation changes} (\#5, 2.0\%): 
     annotation changes specifically refer to the motivation to modify the content of annotations. Annotations are descriptions of code objects, and their main purpose is to increase the readability of the code, so the ``Why'' for comment changes is common sense. For example, \textit{``add docs about null responses''} [\#261].
    
    \item[\textsc{mw5}] \textbf{Code refactoring} (\#15, 6.0\%): 
    these changes involve refactoring and formatting of code. Changes may include polishing, formatting, renaming, cleaning up and other similar operations to improve the readability of the code. 
    For example, \textit{``Polish pom.xml. Apply consistent formatting, drop JDK 8 support and cleanup repo [...]''} [\#S47].
    
    \item[\textsc{mw6}] \textbf{Version management} (\#5, 2.0\%): 
    these commits include changes that involve version management, such as the updating of version numbers. This is an essential step as it tags a specific version of the software, which is important for maintainability, e,g., \textit{``prepare version 2.8.1''} [\#S150].
    
\end{itemize}

\subsubsection{What Expression Categories}\label{sec: what patterns}
We identified four categories to express change, i.e. how to express ``What'' in commit messages. We introduce these categories with examples as follows, including their counts and percentages in the 252 analyzed commit messages (indicated in parentheses).
\paragraph{\textbf{Summarize Code Object Change}}
The first category represents commit messages that summarize the changes; effectively a summary of the \texttt{diffs}. 
We identified the following subcategories:

\begin{itemize}
    \item[\textsc{sc1}] \textbf{Characteristics of changes} (\#13, 5.2\%): this subcategory highlights the characteristics of the current code change and compares them with other alternative implementations to summarize code changes. For example, in this commit message the developer described an \textit{``attempt at a 3rd I/O interface''}, and described the implementation as being \textit{``inspired by InputStream and OutputStream, but using growing buffers instead of byte arrays as the core data container''} [\#S141] (advocating against fixed-size byte arrays).
    
    \item[\textsc{sc2}] \textbf{Object of change} (\#143, 56.8\%): commit messages in this subcategory summarize the changes from the point of view of the code objects. Over half of the commits express What was changed by pointing out the changed object, which refers to the key component of this change, and developers highlight this in the message. These code objects include attributes, methods, classes, packages, and so on. For example, \textit{``remove creation of `fat' jar...''} [\#S157].
    
    \item[\textsc{sc3}] \textbf {Change list} (\#6, 2.4\%): commit messages in this subcategory indicate changes of several code objects, involving one or more source files. 
    For example, \textit{``this commit removes the following deprecated properties: * `server.connection-timeout' * `server.use-forward-headers' [...]''} [\#S154].
    
    \item[\textsc{sc4}] \textbf {Contrast before and after} (\#16, 6.4\%): messages in this subcategory contrast the state of code objects before and after changes. The following is an example of a contrast before/after message: \textit{``rename HeldCertificate.Builder.issuedBy() to signedBy()''} [\#S64].
    
\end{itemize}

\paragraph{\textbf{Describe Implementation Principle}} This category represents commit messages describing technical principles underpinning the changes. The implementation rationale shows the process by which the code executes correctly. For example, \textit{``SslContextBuilder was using InetAddress.getByName(null) [...] On Android, null returns IPv6 loopback, which has the name `ip6-localhost' ''} [\#S251]. Only six commits (out of 252, 2.4\%) fell in this category.

\paragraph{\textbf{Illustrate Function}} 
Commit messages in this category summarize and explain code changes from a functional perspective. Unlike describing specific changes in code, these messages pay more attention to functional changes. Such messages inform other contributors what has changed by describing any new behaviors  
introduced by these changes. For example, \textit{``Rename preferred-mapper property so its clear it only applies to JSON''}  [\#S169]. This category is common, 65 out of the 252 analyzed commits express what was changed by illustrating function. 

\paragraph{\textbf{Missing What}}
This category refers to commit messages that lack any specification of what was changed. Typically any such changes are small and simple that can be easily inferred. We find this category in 19 commits (accounting for 7.5\%). 
Common examples are the correction of typographic errors, renaming of source code objects, and adding and removing spaces. 

\subsubsection{Linking Maintenance Dimensions to Commit Messages}\label{sec: link}
As discussed in Sec. \ref{sec: relatedwork}, the nature of maintenance activities varies by type as defined by Mockus and Votta \cite{mockus2000identifying}. Different types of maintenance activities (as per the typology of Mockus and Votta \cite{mockus2000identifying}) tend to take different ways to describe changes. We analyzed the distribution of the expression categories of Why and What in the different development activities (see Table~\ref{expressionDestribution}). It is worth noting that some developers use multiple (but no more than two) expression categories together when describing Why or What, and these messages account for only a small percentage, i.e., no more than 9\% of the 252 messages.

\newcommand\mc[1]{\multicolumn{1}{c}{#1}}

\begin{table}[!t]
\caption{Expression categories across  maintenance activities}

\newcommand{\STAB}[1]{\begin{tabular}{@{}c@{}}#1\end{tabular}}

\robustify\bfseries
\sisetup{
    group-digits=true,
    group-minimum-digits=4,
    table-format=2.1,
    mode=text,
    detect-weight=true, 
    detect-family=true,
    round-mode=places, 
    round-precision=1
}
    \begin{tabular}{p{0.15cm}p{2.8cm}S[table-format=2.1]S[table-format=2.1]S[table-format=2.1]} 
    \toprule 
     & \multirow{2}*{Category} & \mc{Corrective} & \mc{Adaptive} & \mc{Perfective} \\  
     & & \mc{(\#116)} & \mc{(\#63)} & \mc{(\#73)} \\
    \midrule 
    
     \multirow{12}{*}{\STAB{\rotatebox[origin=c]{90}{How to express ``Why''}}}
    & Describe issue & \bfseries 45.69\%  & 12.7\% & 6.9\% \cr
    & \hangindent=1em
    Illustrate requirement & 12.1\% & 22.2\% & 21.9\% \cr 
   & Describe objective & 6.9\% & 7.9\% & 11.0\% \cr
   & Imply necessity & 19.0\% & \bfseries 39.7\% & 26.0\% \cr 
   & Missing Why & 12.1\% & 15.9\% & \bfseries 34.2\% \cr
     \addlinespace
   & \hangindent=1em
Describe issue \& Describe objective & 0.8\% & 0\% & 0\% \cr
   & \hangindent=1em
Describe issue \& Imply necessity & 2.6\% & 0\% & 0\% \cr
   & \hangindent=1em
Illustrate requirement \& Imply necessity & 0.8\% & 1.6\% & 0\% \cr
   \addlinespace
   & \hangindent=1em
   Total & 100.0\% & 100.0\% & 100.0\% \cr
    \midrule
    
    \multirow{10}{*}{\STAB{\rotatebox[origin=c]{90}{How to express ``What''}}}
    & \hangindent=1em Summarize code object Change & \bfseries 58.62\% & \bfseries 60.32\% & \bfseries 76.71\% \cr
    & Illustrate function & 22.41\% & 26.98\% & 8.22\% \cr
    & \hangindent=1em Describe implementation principle & 4.31\% & 1.59\% & 0\%\cr
    & Missing What & 6.1\% & 3.17\% & 13.70\% \cr
    \addlinespace
    & \hangindent=1em
 Summarize code object change \& Illustrate function & 8.62\% & 7.94\% & 1.37\%\cr
 \addlinespace
 & \hangindent=1em
   Total & 100.0\% & 100.0\% & 100.0\%\cr
    \bottomrule
    
    \end{tabular} 
\label{expressionDestribution}
\end{table}

Corrective changes are performed to fix defects in an existing codebase. As shown in Table \ref{expressionDestribution}, this matches our finding that \texttt{Describe Issue} is the most common expression category to explain the reason for corrective changes, with the highest proportion of 45.7\%. Developers also express the rationale of code change by combining \texttt{Describe Issue} with \texttt{Describe objective} (0.8\%) and \texttt{Imply necessity} (2.6\%). Likewise, \texttt{Summary of code object changes} is the most common category to describe the What for this maintenance type in a commit message, at 58.6\%.

For Adaptive changes, developers often describe Why by indirectly implying the necessity of the change (category \texttt{Imply Necessity}), with the highest proportion of 39.7\%, and followed by the \texttt{Illustrate Requirement} category, which clarifies the requirements that underpin the change. A possible reason might be that a new feature or change in the processing or data environment usually indicates a need for a change when it is the first commit. However, the implementation of most new features requires multiple changes, and a developer can explain the motivation of a change by describing the relationship with a feature or change what was accepted previously. Another reason may be the description of a change's improvements or benefits to support adding new features.

Perfective changes that developers make to improve, for example, code readability and quality, usually involve only text files, comments, or tests. 
The motivation for these changes tends to be common sense, so Why information is frequently omitted in the messages (category \texttt{Missing Why}), with a proportion of 34.2\%. 
For other non-functional properties, developers tend to use the \texttt{Imply necessity} and \texttt{Illustrate requirement} categories.

We can see that the most common way of expressing Why varies with the types of maintenance activities, i.e., \texttt{Describe Issue} for corrective changes, \texttt{Imply necessity} for adaptive changes, and \texttt{Missing Why} for perfective changes. However, developers tend to summarize code changes directly when describing changes. In addition to perfective maintenance activities, summarizing the changes from the perspective of functional changes is also a common way for developers. The possible reason is that the improvement of non-functional properties is not reflected in the functionality offered by the software. The \texttt{Missing What} category is unusual for all three maintenance activities.
\vspace{2mm}
\begin{summarybox}
 \textbf{Summary for RQ2:} We identified five expression categories of ``Why'' and four expression categories of ``What.'' Further, we found that developers have different expression preferences when writing commit messages for different activities. The results can help developers write a good commit message.
\end{summarybox}

\subsection{Automatically Classifying Good Messages}\label{sec: rq3}
We performed a ten-fold cross-validation to estimate the classifiers' performance. The 1,597 messages were randomly partitioned into ten subsets of similar size. 
The validation had ten rounds; in each round, nine subsets were used to train the model, and the remaining one was used for testing. A different subset was used for testing in each round. We reported the average performance of the ten rounds. 
To investigate the impact of different classification techniques on the performance of our approach, we achieved our classifiers based on eight common classification techniques respectively (see Sec. \ref{sec: model}) and repeated the evaluation on the same dataset. Table \ref{good message evaluation results} presents the performance of \textit{\textbf{C-Why}} and \textit{\textbf{C-What}} using different classification techniques. These results indicate that Bi-LSTM has the best performance on both \textit{\textbf{C-Why}} and \textit{\textbf{C-What}}, with an accuracy of 84.7\% and 91.0\%.
This result is consistent with prior findings \cite{kamath2018comparative, zhou2015c},
namely that deep learning based neural networks are better at processing text classification tasks. 
Therefore, we chose Bi-LSTM to build our classifiers.

\begin{table}[!h]
\caption{Ten-fold cross validation of message classification techniques}
\robustify\bfseries 
    \begin{tabular}{lcc}
    \toprule
    \multirow{1}{*}{Techniques} & Accuracy C-Why & Accuracy C-What \\ 

    \midrule 
    Bi-LSTM &  \bfseries 84.7\% & \bfseries 91.0\% \\
    
    LSTM & 83.6\% & 90.1\%  \\

    Logistic Regression & 79.1\% & 85.5\%  \\ 

    MLP & 80.3\% & 85.1\%  \\ 

    Random Forest & 76.5\% & 86.1\% \\ 

    Gradient Boosting & 72.5\% & 77.1\%  \\ 

    KNN & 74.5\% & 70.7\%  \\ 

    Decision Tree & 68.5\% & 76.8\% \\ 
    
    \bottomrule 
    \end{tabular} 
    
    






    
    
\label{good message evaluation results}
\end{table}

Table~\ref{BiLSTMperformance} shows that the three Bi-LSTM based classifiers perform very well, with an accuracy of 84.7\%, 91.0\%, and 75.9\%, respectively. Specifically, when determining whether a message misses ``Why'' (\textit{Positive}: missing Why, \textit{Negative}: having Why), our classifier \textit{\textbf{C-Why}} exhibits good performance with a precision of 76.5\% and a recall of 70.9\%. In determining whether a message misses ``What'' (\textit{Positive}: missing What, \textit{Negative}: having What), our classifier \textit{\textbf{C-What}} shows good performance with a precision of 78.2\% and a recall of 64.5\%. According to the output of our classifiers, developers will get a hint of what information is currently missing, so they can review and revise their commit messages.
Further, our classifier \textit{\textbf{C-Good}} also exhibits good performance when identifying well-written messages (\textit{Positive}: well written, \textit{Negative}: needs improvement), with a precision of 81.6\% and a recall of 74.0\%.
Compared to the unfiltered dataset with an average of only 56\% good commit messages, our classifier can automatically identify and construct a higher-quality commit message dataset.

\begin{table}[!t]
\caption{Performance of Bi-LSTM based classifier (ten-fold cross-validation)}

\robustify\bfseries
\sisetup{
    group-digits=true,
    group-minimum-digits=4,
    mode=text,
    detect-weight=true, 
    detect-family=true,
    round-mode=places, 
    round-precision=1
}
\begin{tabular}{llS[table-format=2.1]<{\%}S[table-format=2.1]<{\%}S[table-format=2.1]<{\%}}
  

 
    
    
    
    
    \toprule 
     \multicolumn{2}{c}{Metrics} & \mc{C-Why} & \mc{C-What} & \mc{C-Good} \\    
    \midrule
    \multirow{3}{*}{Positive}
    & Precision & 76.46 &  78.21 & 81.6 \cr

    & Recall & 70.91 & 64.45 & 74.0\cr 
 
    & F1 & 73.08 & 68.89 & 77.6\cr
    \midrule
    
    \multirow{3}{*}{Negative}
    
    & Precision & 88.12 & 93.41 & 70.0 \cr
    
    & Recall  & 90.16 & 96.20 & 78.4 \cr
    
    & F1  & 89.08 & 94.72 & 73.9\cr
    \midrule
    & Accuracy & 84.69 & 91.02 & 75.9\cr
    \bottomrule
\end{tabular}
\label{BiLSTMperformance}
\end{table}

\begin{summarybox}
\textbf{Summary for RQ3:} We proposed three classification models based on \textit{Bi-LSTM} to automatically identify whether a commit message is well-written and whether a commit message contains ``Why'' or ``What.'' All of them performed well in our dataset and can be reused.
\end{summarybox}

\section{Implications}\label{sec: discussion}
Commit messages are of pivotal importance to facilitate coordination and communication among developers and thus it is important to understand what constitutes good messages and how they are written.
We discuss implications for developers and researchers.

\textbf{Implications for Developers}. 
Our analysis of how the Why and What information is expressed offers developers an understanding of what constitutes a good commit message. At the same time, the results of linking maintenance activities to the message expression categories can prompt developers to write better commit messages.
For example, when performing a corrective task, developers could initially choose the most common expression category (applicable in many scenarios), i.e., \texttt{Describe Issue}; this would not only improve the commit message, but may also inspire developers to become more aware of different ways to express why changes are needed. More specifically, the subcategory \texttt{Describe error scenario} can inspire the developer to describe the bug reproduction steps and the background of the change. At the same time, developers can choose the most popular way, i.e., \texttt{Object of change} 
to summarize code changes and describe the changes to key code objects so that other developers can grasp the focus more quickly.
We hope these can improve the 
quality of commit messages in the long term.

We also designed two automatic tools for developers to check whether the commit message being written conforms to a good one, i.e., containing both Why and What. 
With the help of these tools, the developer can know what is missing in his/her commit message and supplement it accordingly.

\textbf{Implications for Researchers}. 
Our study demonstrates that considerable proportions of commit messages are of poor quality (an average of 44\% in the five popular OSS projects), suggesting that this important modality for developers to share knowledge is not used optimally. While this has consequences for the long-term maintainability of any project, we also observed that approaches for automatic generation of commit messages are trained with uncurated data, i.e., datasets with commit messages that are not filtered based on quality. In turn, such models generate low-quality commit messages as well, which exacerbates the issue of poorly written commit messages. 

Nevertheless, it is time-consuming for researchers to curate good commit messages manually. We proposed an automatic classifier (i.e., \textit{\textbf{C-Good}}) that performs well in identifying good commit messages (precision: 81.6\%). This classifier has the potential to help researchers construct a large dataset of high-quality commit messages from massive historical data stored in GitHub and other VCSs. 
In the future, it is necessary to construct a standard dataset of commit messages to facilitate comparison among different generation methods and promote the quality of generated messages.

\section{Threats to Validity}\label{sec: limitation}
We are aware of some threats to validity which we discuss next.
When constructing the dataset, we removed commit messages generated by known bots \cite{erlenhov2019current, dey2020exploratory, Amreen2020, dey2020detecting}. It may be that new bots are becoming more advanced and generating more human-like messages, suggesting a threat where the filtered data may still contain some non-human-written messages. Future work should consider such developments of new bots as they may require more careful data filtering. Notably, among the 1,649 manually analyzed messages, we did not find any new bot patterns, 
which suggest the results of our qualitative analysis were not affected.

Manual labeling of commit messages poses a subjective threat to validity. To minimize this threat, two authors labeled commit messages independently and introduced an experienced colleague in qualitative research to reach an agreement through several discussions. The agreement level (0.91) is high, indicating a high level of reliability. 
Some of the labeled messages were subsequently confirmed by experienced OSS contributors. Further, the manually analyzed messages come from 339 developers. It means that multiple commit messages may originate from the same authors, which pose a threat of limiting the diversity of the message taxonomy. One strand of future work can extend this analysis to include more projects (and written in other languages than Java), to verify and, if needed, enrich the taxonomy of commit messages reported in this study.

Another threat is that the dataset (1,597 messages in total) is not large enough for training classifiers. The dataset was randomly sampled from the five OSS projects, and its limited scale is because of the complexity involved in manual labeling and analysis of the commit messages. To reduce this threat, we used a ten-fold cross-validation procedure to get as much valid information (nine-fold) from the dataset as possible and use the average performance on different test datasets to improve the models' capability to generalize \cite{SantosSAAS18}. However, developers may write better messages over time. The ten-fold cross-validation, i.e., randomly divided ten subsets, did not consider the influence of time on developers' experience in writing message. This is one potential strand for future work to further add rigor to these results. 

When automatically classifying commit messages, other factors may be related to the quality of commit messages. For example, Chahal and Saini \cite{chahal2018developer} proposed 11 format-related metrics to measure the syntax quality of commit messages. Our classifiers only used the text content of messages as input for training. To alleviate this threat, we replaced those format-related elements with a unified token during preprocessing (see Sec.~\ref{sec: model}), such as using ``<enter>'' to represent line brakes. This preprocessing ensures that syntactic features are considered during the classification of commit messages.

Threats to external validity consider the generalization of our findings. The dataset we analyzed was collected from five popular projects on GitHub implemented in Java, thus posing a threat to external validity. 
Our findings may not be generalizable to other projects, whether they are open or closed-source, or projects that use other languages. 
It is very well possible that developers using other programming languages have different message-written patterns that have not been explored in the scope of this work. 
Future studies could investigate more diverse projects to gain a deeper understanding of what constitutes a good commit message. Notwithstanding these limitations, the automatic classification tools we proposed can be easily adapted to other projects with other languages by simply replacing datasets.

\section{Conclusion}\label{sec: conclusion}
Commit messages play an important role in collaborating software development and evolution. Nonetheless, the considerable proportions of low-quality messages in OSS projects reflect the difficulties that developers face when writing commit messages, and threaten the effectiveness of existing automatic commit message generation tools. Our study explored the distribution and expression patterns of these well-written messages, linked message expression categories to different maintenance activities, and construct several automatic identification models of good commit messages. 
Our study findings can help developers write good commit messages and assist researchers to construct high-quality datasets before generating messages automatically. 

\section*{Acknowledgments}
We are grateful to the software engineers who participated in the survey.
This work was sponsored by the National Natural Science Foundation of China (62141209, 61690205, 62172037, and 61772071), and Science Foundation Ireland grants 15/SIRG/3293 and 13/RC/2094-P2.

\newpage
\balance
\bibliographystyle{ACM-Reference-Format}
\bibliography{reference.bib}

\end{document}